\documentclass[preprint,12pt]{elsarticle}
\usepackage{amssymb}
\usepackage{amsmath}

\journal{Physics Letters A}

\begin{document}

\begin{frontmatter}

\title{On the quantum mechanical derivation of the Wallis formula for $\pi$}

\author{O.~I.~Chashchina}
\address{\'{E}cole Polytechnique, Palaiseau, France}
\ead{chashchina.olga@gmail.com}

\author{ Z.~K.~Silagadze}
\address{Budker Institute of Nuclear Physics and
Novosibirsk State University, 630 090, Novosibirsk, Russia}
\ead{silagadze@inp.nsk.su}

\begin{abstract}
We comment on the Friedmann and Hagen's quantum mechanical derivation of 
the Wallis formula for $\pi$. In particular, we demonstrate that not only the 
Gaussian trial function, used by Friedmann and Hagen, but also the Lorentz 
trial function can be used to get the Wallis formula. The anatomy of the 
integrals leading to the appearance of the Wallis ratio is carefully revealed. 
\end{abstract}


\end{frontmatter}

\section{Introduction}
Recently Friedmann and Hagen presented an interesting quantum mechanical 
derivation of the Wallis formula for $\pi$ \cite{1}. Their result follows from
comparing  the variational principle estimate of the hydrogen atom lowest 
energy level for a given orbital angular momentum quantum number with the 
exact quantum mechanical result. The Bohr's correspondence principle is used 
to argue that in the limit of large angular momentum quantum numbers the two 
results should coincide.

Friedmann and Hagen use the Gaussian trial function in their variational 
calculation. So a question naturally arises whether their recovery of the 
Wallis formula from these calculations is conditioned by the use of this 
particular form of the trial function. The answer is no because, as we 
demonstrate below, the Lorentz trial function also enables us to get the 
Wallis formula for $\pi$ from the variational approach to the spectrum of the 
hydrogen atom.

The paper is organized as follows. For reader's convenience, we first 
reproduce the main steps of the Friedmann and Hagen's approach. Then we prove 
an interesting identity for the sum of the corresponding  variational energy 
levels and show how to generalize this identity. The Lorentz trial function is 
considered in the next section, and we conclude that it also leads to the
Wallis formula for $\pi$. Then we try somewhat to demystify the Friedmann and 
Hagen's proof of the Wallis formula. Finally, we provide some concluding 
remarks.

\section{Friedmann and Hagen's approach}
The Hamiltonian for the hydrogen atom has the form
\begin{equation}
\hat H=-\frac{\hbar^2}{2mr^2}\left[\frac{\partial}{\partial r}\left(r^2
\frac{\partial}{\partial r}\right)-\hat l^2\right]-\frac{e^2}{r},
\label{eq1}
\end{equation}
where
\begin{equation}
\hat l^2=-\frac{1}{\sin^2{\theta}}\left[\sin{\theta}\frac{\partial}
{\partial \theta}\left(\sin{\theta}\frac{\partial}{\partial \theta}\right)+
\frac{\partial^2}{\partial \phi^2}\right]
\label{eq2}
\end{equation}
is the square of the angular momentum operator (divided by $\hbar^2$) in 
spherical coordinates. 

For variational estimation of the hydrogen atom energy Friedmann and Hagen
use the Gaussian trial function
\begin{equation}
\Psi_{\alpha lm}=r^le^{-\alpha r^2}Y_l^m(\theta,\phi),
\label{eq3}
\end{equation}
where $Y_l^m(\theta,\phi)$ are the usual spherical harmonics satisfying
\begin{equation}
\hat l^2 \,Y_l^m(\theta,\phi)=l(l+1)Y_l^m(\theta,\phi).
\label{eq4}
\end{equation}
This trial function has no nodes and thus corresponds to the zero radial 
quantum number.

Using the orthonormality relation for spherical harmonics and the integral
\begin{equation}
I_m=\int\limits_0^\infty x^me^{-x^2}=\frac{1}{2}\,\Gamma\left(\frac{m+1}{2}
\right),
\label{eq5}
\end{equation}
it is straightforward to get an expectation value of the Hamiltonian 
(\ref{eq1}) in the quantum state (\ref{eq3}). The result is
\begin{equation}
<\hat H>=\frac{<\Psi_{\alpha lm}|\hat H|\Psi_{\alpha lm}>}{<\Psi_{\alpha lm}|
\Psi_{\alpha lm}>}=\frac{\hbar^2\alpha}{m}\left(l+\frac{3}{2}\right)-
e^2\sqrt{2\alpha}\,\frac{\Gamma(l+1)}{\Gamma\left (l+\frac{3}{2}\right)}.
\label{eq6}
\end{equation}
Minimizing this expectation value with respect to $\alpha$, we get the 
following variational energy levels of the hydrogen atom:
\begin{equation}
E^{(\alpha)}_l=-\frac{me^4}{2\hbar^2}\,\frac{1}{l+\frac{3}{2}}\left[
\frac{\Gamma(l+1)}{\Gamma\left(l+\frac{3}{2}\right)}\right]^2=
-\frac{me^4}{2\hbar^2}\,\frac{1}{n+\frac{1}{2}}\left[
\frac{\Gamma(n)}{\Gamma\left(n+\frac{1}{2}\right)}\right]^2,
\label{eq7}
\end{equation}
with $n=l+1$.

It is expected \cite{1} from Bohr's correspondence principle that for large 
orbital momentum quantum numbers $l$ the approximate energy levels (\ref{eq7}) 
should approach the exact quantum mechanical result
\begin{equation}
E_{0,l}=-\frac{me^4}{2\hbar^2}\,\frac{1}{(l+1)^2}=
-\frac{me^4}{2\hbar^2}\,\frac{1}{n^2}.
\label{eq8}
\end{equation} 
Therefore,
\begin{equation}
\lim_{n\to\infty}\frac{n^2}{n+\frac{1}{2}}\left[\frac{\Gamma(n)}
{\Gamma\left(n+\frac{1}{2}\right)}\right]^2=\lim_{n\to\infty}\frac{1}
{n+\frac{1}{2}}\left[\frac{\Gamma(n+1)}{\Gamma\left(n+\frac{1}{2}\right)}
\right]^2=1.
\label{eq9}
\end{equation}
Due to the relations $\Gamma(x+1)=x\Gamma(x)$ and $\Gamma(1/2)=\sqrt{\pi}$, 
we have
\begin{eqnarray} &&
\frac{1}{n+\frac{1}{2}}\left[\frac{\Gamma(n+1)}{\Gamma\left(n+\frac{1}{2}
\right)}\right]^2=\frac{1}{2n+1}\left[\frac{1\cdot 2\cdot 3\cdots n}{
\frac{2n-1}{2}\cdot\frac{2n-3}{2}\cdots\frac{3}{2}}\right]^2\frac{2}{\pi}=
\nonumber \\ &&
\frac{2}{\pi}\,\frac{2\cdot 2}{1\cdot 3}\,\frac{4\cdot 4}{3\cdot 5}\cdots
\frac{2(n-1)\cdot 2(n-1)}{(2n-3)\cdot (2n-1)}\,\frac{2n\cdot 2n}{(2n-1)
\cdot 2(n+1)}.
\label{eq10}
\end{eqnarray}
Hence, the limit (\ref{eq9}) motivated by the quantum physics of the hydrogen 
atom implies the Wallis formula for $\pi$:
\begin{equation}
\frac{\pi}{2}=\frac{2\cdot 2}{1\cdot 3}\,\frac{4\cdot 4}{3\cdot 5}\,
\frac{6\cdot 6}{5\cdot 7}\cdots=\prod_{n=1}^\infty\frac{(2n)^2}{(2n-1)
(2n+1)}.
\label{eq11}
\end{equation}

\section{Some interesting identities}
There is still an other interesting aspect of the approximate energy levels 
(\ref{eq7}). Having in mind that the exact energy levels (\ref{eq8}) resonate
with the celebrated Euler result
\begin{equation}
\sum_{n=1}^\infty\frac{1}{n^2}=\frac{\pi^2}{6},
\label{eq12}
\end{equation}
one may ask if the analogous simple expression can be obtained for the sum
of the series
\begin{equation*}
\sum_{n=1}^\infty\frac{1}{n+\frac{1}{2}}\left[
\frac{\Gamma(n)}{\Gamma\left(n+\frac{1}{2}\right)}\right]^2.
\end{equation*}
This is really possible! Let us denote
\begin{equation}
a_n=\frac{1}{n+\frac{1}{2}}\left[\frac{\Gamma(n)}{\Gamma\left(n+\frac{1}{2}
\right)}\right]^2.
\label{eq13}
\end{equation}
Then 
\begin{equation}
\frac{a_n}{a_{n-1}}=\frac{(n-1)^2}{n^2-\frac{1}{4}}=\frac{4(n-1)^2}
{4n^2-1},
\label{eq14}
\end{equation}
and
\begin{equation}
4n^2a_n=4(n-1)^2a_{n-1}+a_n.
\label{eq15}
\end{equation}
Applying this recurrence relation again and again, we get
\begin{equation}
4n^2a_n=4a_1+a_2+a_3+\ldots+a_n.
\label{eq16}
\end{equation}
Therefore,
\begin{equation}
s_n=\sum\limits_{i=1}^na_i=4n^2a_n-3a_1.
\label{eq17}
\end{equation}
As 
\begin{equation}
\lim\limits_{n\to\infty}n^2a_n=\lim\limits_{n\to\infty}\frac{1}
{n+\frac{1}{2}}\left[\frac{\Gamma(n+1)}{\Gamma\left(n+\frac{1}{2}\right)}
\right]^2=1,
\label{eq18}
\end{equation}
and $a_1=\frac{8}{3\pi}$,
we obtain 
\begin{equation}
\sum_{n=1}^\infty\frac{1}{n+\frac{1}{2}}\left[
\frac{\Gamma(n)}{\Gamma\left(n+\frac{1}{2}\right)}\right]^2=
\lim\limits_{n\to\infty}s_n=4-\frac{8}{\pi}.
\label{eq19}
\end{equation}
This result can be generalized even further. Let
\begin{equation}
b_n=\frac{\Gamma(n+m)\Gamma(n+k)}{\Gamma\left(n+m+\frac{1}{2}\right)\Gamma
\left(n+k+\frac{3}{2}\right)}.
\label{eq20}
\end{equation}
Then
\begin{equation}
\frac{b_n}{b_{n-1}}=\frac{(n+m-1)(n+k-1)}{(n+m)(n+k)+\frac{1}{2}(m-k)-
\frac{1}{4}},
\label{eq21}
\end{equation}
and
\begin{equation}
\frac{4(n+m)(n+k)}{2(k-m)+1}\,b_n=\frac{4(n-1+m)(n-1+k)}{2(k-m)+1}\,b_{n-1}+
b_n.
\label{eq22}
\end{equation}
Similarly to the recurrence relation (\ref{eq15}), (\ref{eq22}) implies
\begin{equation}
\sum\limits_{i=1}^nb_i=\frac{4(n+m)(n+k)}{2(k-m)+1}\,b_n-\left[
\frac{4(m+1)(k+1)}{2(k-m)+1}-1\right]b_1.
\label{eq23}
\end{equation}
But due to (\ref{eq9})
\begin{equation} 
\lim\limits_{n\to\infty}(n+m)(n+k)b_n=
\lim\limits_{n\to\infty}
\sqrt{\frac{n+m+\frac{1}{2}}
{n+k+\frac{1}{2}}}\,A(n+m)A(n+k)=1,
\label{eq24}
\end{equation}
where
$$A(n)=\frac{1}{\sqrt{n+\frac{1}{2}}}\,\frac{\Gamma(n+1)}
{\Gamma\left(n+\frac{1}{2}\right)}.$$
Therefore, taking the limit $n\to\infty$ in (\ref{eq23}), we get
\begin{eqnarray} &&
\sum\limits_{n=1}^\infty \frac{\Gamma(n+m)\Gamma(n+k)}
{\Gamma\left(n+m+\frac{1}{2}\right)\Gamma\left(n+k+\frac{3}{2}\right)}=
\nonumber \\ &&
\frac{4}{2(k-m)+1}\left[1-\frac{\Gamma(m+1)\Gamma(k+1)}{\Gamma\left(m+\frac{1}
{2}\right)\Gamma\left(k+\frac{3}{2}\right)}\right],
\label{eq25}
\end{eqnarray}
where we have simplified the term containing $b_1$ by using 
$4(m+1)(k+1)-2(k-m)-1=4\left(m+\frac{1}{2}\right)\left(k+\frac{3}{2}\right)$, 
and the $\Gamma(x+1)=x\Gamma(x)$ property of the gamma function.
When $m=k=0$, this more general relation is reduced to (\ref{eq19}). 
In fact, (\ref{eq25}) is valid for any real $m$ and $k$.
 
As we see, not only the Wallis formula is hidden in the quantum-mechani\-cal 
hydrogen atom, but also the identity (\ref{eq25}) which Euler probably had
missed to prove, although the Euler result (\ref{eq12}) is not contained in 
it.

\section{Lorentz trial function}
A pedagogical usefulness of the Lorentz trial function
\begin{equation}
\Phi(r)=\frac{1}{a^2+r^2}
\label{eq26}
\end{equation}
was emphasized in \cite{2}. Let us generalize (\ref{eq26}) as follows
\begin{equation}
\Psi_{alm}=\frac{r^l}{(a^2+r^2)^{l+1}}Y_l^m(\theta,\phi),
\label{eq27}
\end{equation}
and see what will change in the Friedmann and Hagen's derivation.

Using the integral
\begin{equation}
I_{m,n}=\int\limits_0^\infty\frac{x^m}{(1+x^2)^n}\,dx=\frac{1}{2}\,
\frac{\Gamma\left(\frac{m+1}{2}\right)\Gamma\left(n-\frac{m+1}{2}\right)}
{\Gamma(n)},
\label{eq28}
\end{equation}
which can be obtained from
\begin{equation}
\int\limits_0^{\pi/2}\sin^{2p-1}(\theta)\,\cos^{2q-1}(\theta)\,d\theta=
\frac{1}{2}\,\frac{\Gamma(p)\Gamma(q)}{\Gamma(p+q)},
\label{eq29}
\end{equation}
by substitution $x=\tan{\theta}$, we can easily find the expectation value
of the Hamiltonian (\ref{eq1}) with respect to this new trial function:
\begin{equation} 
<\hat H>=\frac{\hbar^2}{2m}\,(l+1)\left(l+\frac{1}{2}\right)\,\frac{1}{a^2}-
\frac{e^2}{a}\,\frac{1}{l+\frac{1}{2}}\left[\frac{\Gamma(l+1)}
{\Gamma\left(l+\frac{1}{2}\right)}\right]^2.
\label{eq30}
\end{equation}
Minimizing it with respect to $a$, we get the variational energy levels of the 
hydrogen atom corresponding to the Lorentz trial function:
\begin{equation}
E^{(a)}_l=-\frac{me^4}{2\hbar^2}\,\frac{1}{(l+1)\left(l+\frac{1}{2}\right)^3}
\left[\frac{\Gamma(l+1)}{\Gamma\left(l+\frac{1}{2}\right)}\right]^4=
-\frac{me^4}{2\hbar^2}\,\frac{n-\frac{1}{2}}{n}\left[
\frac{\Gamma(n)}{\Gamma\left(n+\frac{1}{2}\right)}\right]^4.
\label{eq31}
\end{equation}
Therefore, we expect
\begin{equation}
\lim_{l\to\infty}\frac{E^{(a)}_l}{E_{0,l}}=\lim_{n\to\infty}
\frac{n-\frac{1}{2}}{n^3}\left[\frac{\Gamma(n+1)}{\Gamma\left(n+\frac{1}{2}
\right)}\right]^4=1.
\label{eq32}
\end{equation}
But (\ref{eq32}) and (\ref{eq9}) are equivalent, because
$$\frac{n-\frac{1}{2}}{n^3}\left[\frac{\Gamma(n+1)}{\Gamma\left(n+\frac{1}{2}
\right)}\right]^4=\frac{\left(n-\frac{1}{2}\right)\left(n+\frac{1}{2}\right)^2}
{n^3}\left[\frac{1}{n+\frac{1}{2}}\left(\frac{\Gamma(n+1)}{\Gamma\left(
n+\frac{1}{2}\right)}\right)^2\right]^2.$$

As we see, the Lorentz trial function also can be used to get a quantum 
mechanical proof of the Wallis formula. However, unlike (\ref{eq19}), it 
doesn't seem possible to get a closed form for the sum of approximate energy 
eigenvalues (\ref{eq31}). Just for completeness, let us mention that we can 
calculate approximately this sum by using the following relation \cite{3}:
\begin{equation}
\sqrt[4]{x^2+\frac{1}{2}x+\frac{1}{8}-\frac{1}{128x}}<\frac{\Gamma(x+1)}
{\Gamma\left(x+\frac{1}{2}\right)}<\sqrt[4]{x^2+\frac{1}{2}x+\frac{1}{8}}.
\label{eq33}
\end{equation}

If we apply the above described procedures to the isotropic quantum harmonic 
oscillator with the potential energy $V(r)=\frac{1}{2}m\omega^2r^2$, nothing
notably happens. The exact energy levels (for zero radial quantum number) 
$E_{0,l}=\hbar\omega\left(l+\frac{3}{2}\right)$ are reproduced, if we use the 
Gaussian trial function. Whereas the Lorentz trial function gives the 
approximate energy levels
$$E^{(a)}_l=\hbar\omega\sqrt{\frac{(l+1)\left(l+\frac{1}{2}\right) 
\left(l+\frac{3}{2}\right)}{l-\frac{1}{2}}},$$
and in this case we obtain the valid but uninteresting limit
$$\lim_{l\to\infty}\frac{(l+1)\left(l+\frac{1}{2}\right)}
{\left(l+\frac{3}{2}\right)\left(l-\frac{1}{2}\right)}=1.$$

\section{Magic demystified}
At first sight it may appear magical that a classic formula for $\pi$ is 
hidden inside the quantum mechanics of the hydrogen atom. Moreover, two very 
different trial functions both lead to the very same formula.
However, on second thought, it becomes clear that the appearance of the 
Wallis formula in the variational treatment of the hydrogen atom is related
to the presence of the Wallis ratio \cite{4}
\begin{equation}
W_n=\frac{(2n-1)!!}{(2n)!!}=\frac{1}{\sqrt{\pi}}\,\frac{\Gamma\left(n+
\frac{1}{2}\right)}{\Gamma(n+1)}  
\label{eq35}
\end{equation}
in the relevant integrals, or in the ratios of the relevant integrals.

In case of the Gaussian trial function, the primary source for the Wallis 
ratio appearance is the integral (\ref{eq5}). In particular, the following
ratio of the integrals  (\ref{eq5}) emerges while calculating the expectation
 value of the potential energy:
\begin{equation}
\frac{I_{2l+1}}{I_{2l+2}}=\frac{\Gamma(l+1)}{\Gamma\left(l+\frac{3}{2}
\right)}=\frac{\Gamma(n)}{\Gamma\left(n+\frac{1}{2}\right)}=\frac{1}
{\sqrt{\pi}nW_n}.
\label{eq36}
\end{equation}
The integral (\ref{eq5}) itself can be understood on the base of the 
recurrence relation
\begin{equation}
I_m=-\frac{1}{2}\int\limits_0^\infty x^{m-1}de^{-x^2}=\frac{1}{2}\int\limits_0
^\infty e^{-x^2}dx^{m-1}=\frac{m-1}{2}\,I_{m-2}.
\label{eq37}
\end{equation}

In case of the Lorentz trial function, we have the following normalization 
integral:
\begin{eqnarray} &&
\hspace*{-20mm}
I_{2l+2,2l+2}=\frac{(-1)^{l+1}}{(l+1)(l+2)\cdots(2l+1)}\left . 
\frac{\partial ^{l+1}}{ \partial\alpha^{l+1}}\int\limits_0^\infty\frac{dx}
{(1+\alpha x^2)^{l+1}}\right|_{\alpha=1}=\nonumber\\ && \hspace*{-20mm}
\frac{(-1)^{l+1}l!}{(2l+1)!}
\int\limits_0^\infty \frac{dx}{(1+x^2)^{l+1}}\left. \frac{\partial ^{l+1}}
{ \partial\alpha^{l+1}}\frac{1}{\sqrt{\alpha}}\right|_{\alpha=1}=\frac{1}
{2^{2l+1}}\int\limits_0^\infty \frac{dx}{(1+x^2)^{l+1}}.
\label{eq40}
\end{eqnarray}
It is well known \cite{5} that the rational integral
\begin{equation}
G_{l+1}=\int\limits_0^\infty \frac{dx}{(1+x^2)^{\,l+1}}
\label{eq41}
\end{equation}
is closely related to the Wallis ratio. Namely, writing
$$\frac{dx}{(1+x^2)^{l+1}}=\frac{(1+x^2-x^2)\,dx}{(1+x^2)^{l+1}}=
\frac{dx}{(1+x^2)^l}+\frac{1}{2l}\,d\left(\frac{1}{(1+x^2)^l}
\right), $$
and integrating the second term by parts, we get a recurrence relation
\begin{equation}
G_{l+1}=\frac{2l-1}{2l}\,G_l=\frac{1}{2^2}\,\frac{2l(2l-1)}{l^2}\,G_l.
\label{eq42}
\end{equation}
Since $G_1=\frac{\pi}{2}$, this recurrence relation implies that \cite{5}
\begin{equation}
G_{l+1}=\frac{1}{2^{2l}}\,\frac{(2l)!}{l!\,l!}\,\frac{\pi}{2}=\frac{1}
{2^{2l}}\,\frac{\Gamma(2l+1)}{\Gamma^(l+1)\Gamma^(l+1)}\,\frac{\pi}{2}=
\frac{\pi}{2}\,W_l,
\label{eq43}
\end{equation}
where the last step follows from the Legendre's duplication formula for the 
gamma function applied to $\Gamma(2l+1)$:
$$\Gamma(2l+1)=\frac{2^{2l}\Gamma(l+1)\Gamma\left(l+\frac{1}{2}\right)}
{\sqrt{\pi}}.$$
Therefore,
\begin{equation}
I_{2l+2,2l+2}=\frac{\pi}{2^{2l+2}}\,W_l.
\label{eq44}
\end{equation}
On the other hand, integral related to the Coulomb potential has the form
\begin{eqnarray} &&
\hspace*{-20mm}
I_{2l+1,2l+2}=\frac{(-1)^l}{(l+2)(l+3)\cdots(2l+1)}\left . 
\frac{\partial ^l}{ \partial\alpha^l}\int\limits_0^\infty\frac{dx}
{(1+\alpha x^2)^{l+2}}\right|_{\alpha=1}=\nonumber\\ && \hspace*{-20mm}
\frac{(-1)^l(l+1)!}{(2l+1)!}
\int\limits_0^\infty \frac{x\,dx}{(1+x^2)^{l+2}}\left. \frac{\partial^l}
{\partial\alpha^l}\frac{1}{\alpha}\right|_{\alpha=1}=\frac{1}{2}\,
\frac{(l!)^2}{(2l+1)!}.
\label{eq45}
\end{eqnarray}
Using again the Legendre's duplication formula applied to $\Gamma(2l+2)$,
we get
\begin{equation}
I_{2l+1,2l+2}=\frac{1}{2^{2l+2}}\,\frac{\sqrt{\pi}\,\Gamma(l+1)}{\Gamma
\left(l+\frac{3}{2}\right)}=
\frac{1}{2^{2l+2}}\,\frac{1}{\left(l+\frac{1}{2}\right)W_l}.
\label{eq46}
\end{equation}
Therefore,
\begin{equation}
\frac{I_{2l+1,2l+2}}{I_{2l+2,2l+2}}=\frac{1}{l+\frac{1}{2}}\,\frac{1}
{\pi W_l^2},
\label{eq47}
\end{equation}
which makes clear the origin of the squared Wallis ratio in the expectation
value of the Coulomb potential energy.

\section{Concluding remarks}
It is remarkable and fascinating that a purely quantum mechanical proof of the
Wallis formula can be given. However, there is no magic here, only certain
easy to understand properties of the integrals involved.

Of course, there is no need in the Bohr's correspondence principle to prove 
(\ref{eq9}). Indeed, Stirling's series can be used to get
$$\frac{\Gamma(z+\alpha)}{\Gamma(z+\beta)}\sim z^{\alpha-\beta},$$
as $z\to\infty$ (see, for example, \cite{6}). Then
$$\lim_{l\to\infty}\frac{(l+1)^2}{l+\frac{3}{2}}\left[\frac{\Gamma(l+1)}
{\Gamma\left(l+\frac{3}{2}\right)}\right]^2=\lim_{l\to\infty}\frac{(l+1)^2}
{\left(l+\frac{3}{2}\right)l}=1$$
follows quite simply.

In fact, (\ref{eq9}) is a particular case of the more general Wendel's
limit \cite{7}
$$\lim\limits_{x\to\infty}\frac{\Gamma(x+s)}{x^s\,\Gamma(x)}=1$$
valid for any real $s$ and $x$.

An other simple way to prove (\ref{eq9}) is provided by Kazarinoff's double
inequality \cite{8}
$$\sqrt{n+\frac{1}{4}}<\frac{\Gamma(n+1)}{\Gamma\left(n+\frac{1}{2}\right)}
<\sqrt{n+\frac{1}{2}}.$$
Generalizations of the  Wendel's and Kazarinoff's results produced a vast 
mathematical literature on bounds for the ratio of two Gamma functions
(see, for example, review papers \cite{9,10,11}).

Of course, these comments don't diminish the charm of Friedmann and Hagen's 
derivation. No matter how simple the purely mathematical proof of (\ref{eq9}) 
may seem, no one suspected to find Wallis formula for $\pi$ in hydrogen atom 
until Friedmann and Hagen revealed this beautiful connection between $\pi$ and 
quantum mechanics. 

\section*{Acknowledgments}
The work is supported by the Ministry of Education and Science of the Russian 
Federation. We would also like to acknowledge very insightful comments 
from R.~Israel \cite{12} and T.~Amdeberhan \cite{13}.

\end{document}